\documentstyle[prb,aps]{revtex}
\begin{document}
\draft
\title{Confinement of two-dimensional excitons in a non-homogeneous magnetic field}
\author{J. A. K. Freire\cite{AddJAKF}, A. Matulis\cite{AddMat}, and F. M. Peeters\cite{AddFMP}}
\address{Departement Natuurkunde, Universiteit Antwerpen (UIA),
Universiteitsplein 1, B-2610 Antwerp, Belgium}
\author{V. N. Freire and G. A. Farias}
\address{Departamento de F\'{\i}sica, Universidade Federal do Cear\'a, Centro de Ci\^encias Exatas,
Campus do Pici, Caixa Postal 6030, 60455-760 Fortaleza, Cear\'a,
Brasil}

\date{\today}
\maketitle
\begin{abstract}
The effective Hamiltonian describing the motion of an exciton in
an external non-homogeneous magnetic field is derived. The
magnetic field plays the role of an effective potential for the
exciton motion, results into an increment of the exciton mass and
modifies the exciton kinetic energy operator. In contrast to the
homogeneous field case, the exciton in a non-homogeneous magnetic
field can also be trapped in the low field region and the field
gradient increases the exciton confinement. The trapping energy
and wave function of the exciton in a GaAs two-dimensional
electron gas for specific circular magnetic field configurations
are calculated. The results show than excitons can be trapped by
non-homogeneous magnetic fields, and that the trapping energy is
strongly correlated with the shape and strength of the
non-homogeneous magnetic field profile.
\end{abstract}


\section{Introduction}

Two dimensional confinement of excitons or atoms by magnetic
fields is an important phenomenon in physics, and is expected to
bring a significant improvement to lasers and other functional
devices~\cite{Mew}. Numerous works have been carried out on the
study of the exciton properties in applied magnetic
fields~\cite{Gor,Ler,Coo,Dzy,Kula}. However, for the most part,
all the attention has been focused on the influence of homogenous
fields on the exciton inner properties. Several theoretical
investigations and experiments show the possibility of trapping
and guiding of atoms by means of non-homogeneous magnetic
fields~\cite{Wolf,Tollett}. Recently, numerous
papers~\cite{FMPrev} have appeared on the properties of charged
particles, like e. g. electrons, in different non-homogeneous
magnetic field profiles. For example, quantum mechanical bound
states were found which had no classical analogue. Christianen
{\it et al}~\cite{FabioF,Fabio} performed photoluminescense (PL)
measurements on excitons in the presence of non-homogeneous
magnetic fields. Strips of ferromagnetic material (see, e.g.,
Ref.~\cite{Geim}) on top of a quantum well were used to create
strong magnetic field gradients. They concluded that excitons are
forced to regions of low field gradient and that magnetic traps
for excitons are feasible. To the best of our knowledge, there is
no theoretical work in the literature on studies of excitons in
non-homogeneous magnetic fields.

The purpose of the present paper is to develop a mathematical
formalism for calculating the quantum mechanical properties of
excitons in a non-homogeneous magnetic field, and to illustrate
the trapping possibilities of the excitons in some magnetic field
profiles. These ones are created by the deposition of magnetic
disks and also superconducting disks on top of a two dimensional
electron gas (2DEG) with a homogeneous field applied perpendicular
to the 2DEG. It results in a non-homogeneous magnetic dipole type
of profile and in a magnetic antidot~\cite{BJ} profile,
respectively, in the 2DEG.

The layout of the paper is as follows. In Sec.~II the effective
Hamiltonian describing the exciton motion in the non-homogeneous
magnetic field is derived. In Sec. III we consider circular
symmetric magnetic field profiles and reduce the exciton
Hamiltonian to a one-dimensional (1D) Schr\"odinger equation
subjected to a spatially dependent effective potential and
effective mass. In Sec. IV we calculate explicitly the magnetic
field profiles used in our work and study how strongly they are
able to trap excitons. Our numerical results for the exciton
trapping energies are discussed in Sec. V, and the conclusions are
given in Sec. VI.

\section{The effective Hamiltonian}
We consider a 2D system of two particles with opposite charge
interacting via the Coulomb interaction and moving in a
non-homogeneous magnetic field characterized by the vector
potential ${\bf A}({\bf r})$. The Hamiltonian of that system is
given by:

\begin{eqnarray}
H=\frac{\hbar ^2}{2m_e}\left\{ -i\nabla _e+\frac{e}{\hbar c} {\bf
A}({\bf r}_e)\right\}^2+ \frac{\hbar ^2}{2 m_h}\left\{ -i\nabla
_h-\frac{e}{\hbar c} {\bf A}({\bf r}_h)\right\}^2
-\frac{e^2}{\varepsilon \left|{\bf r}_e-{\bf r }_h\right|},
\end{eqnarray}
where $\varepsilon$ is the dielectric constant of the material the
exciton is moving in, $m_e$ ($m_h$) is the electron (hole)
effective mass, and ${\bf r_e}$, ${\bf r_h}$ are the electron and
hole coordinates, respectively, in the $xy$-plane. To simplify the above
Hamiltonian, we use the center of mass $ {\bf R} = (m_e {\bf r}_e
+ m_h{\bf r}_h)/M $ and relative motion coordinates $ {\bf r}={\bf
r}_e - {\bf r}_h $, where $ M=m_e+m_h $ is the total mass of the
exciton. Next, in order to obtain the correct asymptotic behavior,
we apply a wave function phase transformation analogous to the one
used by Gor'kov and Dzyaloshinsky\cite{Gor} in the case of a
homogeneous magnetic field:

\begin{equation}
\Psi({\bf R},{\bf r}) \to \exp\left\{ -i(e/\hbar c){\bf r} \cdot
{\bf A} \left({\bf R} \right) \right\} \Psi({\bf R},{\bf r}),
\end{equation}
which leads to the following transformed Hamiltonian:

\begin{eqnarray}
H &=& \frac{\hbar ^2}{2m_e}\left\{ -i\frac{m_e}{M}\nabla
_R-i\nabla_r + \frac{e}{\hbar c} {\bf A}\left({\bf R} +
\frac{m_h}{M}{\bf r}\right)- \frac{e m_e}{\hbar c M}\nabla_R
\{{\bf r A({\bf R})}\} - \frac{e}{\hbar c} {\bf A}({\bf R})
\right\}^2 \nonumber \\ &+& \frac{\hbar ^2}{2m_h}\left\{
-i\frac{m_h}{M}\nabla _R + i\nabla_r - \frac{e}{\hbar c} {\bf
A}\left({\bf R} - \frac{m_e}{M}{\bf r}\right)- \frac{e m_h}{\hbar
c M}\nabla_R \{{\bf r A({\bf R})}\} + \frac{e}{\hbar c} {\bf
A}({\bf R}) \right\}^2 \nonumber \\ &-& \frac{e^2}{\varepsilon r}.
\end{eqnarray}

In real experimental situations~\cite{FMPrev,Fabio} the size of
the exciton is smaller than the length scale over which the
magnetic field varies, which allows us to use the adiabatic
approach~\cite{Coo}. We assume that the exciton relative motion is
fast as compared with its center-of-mass motion, and that the
characteristic dimension in the relative coordinate (i.e. the
exciton radius) is much smaller than that of the center-of-mass
motion (i.e. the characteristic length of the magnetic field
inhomogeneity). This allows us to expand the vector potential
${\bf A}({\bf R}\pm m_{h(e)}{\bf r}/M)$ into $\mathbf{r}$
power series restricting the consideration with up to second
order terms. Note that the exciton radius in GaAs, e.g., is
typically $a_B^* = \varepsilon \hbar^2 / \mu e^2 \simeq
120\thinspace\AA$, while the magnetic field inhomogeneity varies
typically on a micron scale\cite{Fabio,Geim,GeimPt}. Within the
adiabatic approximation, we obtain the following expression for
the exciton Hamiltonian:

\begin{eqnarray}
H &=& \frac{\hbar^2}{2\mu}\Big( -i\nabla_r + \xi\frac{e}{\hbar c}
\{( {\bf r}\cdot\nabla_R) {\bf A}({\bf R})\} \Big)^2 +
\frac{\hbar^2}{2M}\Big( i\nabla_R + \frac{e}{\hbar c}[{\bf
r}\times{\bf B}({\bf R})] \Big)^2 - \frac{e^2}{\varepsilon r},
\label{eq4}
\end{eqnarray}
where  $ \mu = m_e m_h / M $ is the exciton reduced mass and $\xi
= (m_h -m_e)/M $. The first term in the Hamiltonian,(Eq.
\ref{eq4}), describes the kinetic energy of the exciton relative
motion, in which the expression $\xi({\bf r}\cdot\nabla_R){\bf
A}({\bf R})$ can be interpreted as an effective vector potential
describing the local magnetic field $\xi{\bf B}({\bf R)}$. We can
choose the gauge such that $({\bf r}\cdot\nabla_R){\bf A}({\bf
R})={\bf B({\bf R})}\times{\bf r}/2 + \nabla_r\left\{({\bf
r}\cdot\nabla_R)\left\{{\bf r}\cdot{\bf A}({\bf
R})\right\}\right\}/2$, and simplify the second term of this
expression by applying the following transformation:

\begin{equation}
\Psi({\bf R},{\bf r}) \to \exp\left\{ -i\xi(e/\hbar c)\Omega(r)
\right\} \Psi({\bf R} ,{\bf r}),
\end{equation}
with $\Omega(r) = ({\bf r}\cdot\nabla_R)\{{\bf r} \cdot{\bf
A}{(\bf R})\}/2$. The above transformation does not change the
center-of-mass motion, and leads only to the appearance of
$r^3$-order terms in the relative motion, which are neglected in
the present adiabatic approach. In doing so, we obtain the
following transformed Hamiltonian:
\begin{equation}\label{hambas}
H =  -\frac{\hbar^2}{2\mu}\nabla_r^2 - \frac{e^2}{\varepsilon r} +
W_1 + W_2 - \frac{\hbar^2}{2M}\nabla_R^2,
\end{equation}
where terms of first and second power in the magnetic field
strength are denoted by
\begin{eqnarray}
W_{1} &=& \frac{e}{2c\mu} \xi{\bf B}({\bf R}) \cdot {\bf L} +
\frac{ie\hbar}{2M c} \left\{{\bf B({\bf R})}\times\nabla_R -
\nabla_R\times{\bf B({\bf R})}\right\}\cdot{\bf r}, \nonumber\\
W_{2} &=& \frac{e^2}{8c^2\mu} B({\bf R})^2 r^2. \label{W1W2}
\end{eqnarray}

Here the symbol ${\bf L}=\{{\bf r}\times(-i\hbar\nabla_r)\}$
stands for the exciton relative angular momentum operator. Notice
that in case of a homogeneous magnetic field the above Hamiltonian
reduces to the one which can be found in the literature (see Ref.
5 and references therein). The exciton Hamiltonian in a
homogeneous magnetic field consists of a part with no magnetic
field dependence, a part linear (the angular momentum term) in the
magnetic field, and a quadratic part (the diamagnetic shift term)
in the magnetic field. In the present Hamiltonian which describes
the exciton motion in a non-homogeneous magnetic field we also
have these terms, but now they are position dependent. We also
have an additional term with a linear dependence in the magnetic
field (the last term in the $W_{1}$ Hamiltonian, Eq. (\ref{W1W2}))
which results from the gradient in the magnetic field, and which
modifies the exciton mass.

Now within the spirit of the above adiabatic approach, the total
exciton wave function is represented as the product of the wave
functions describing its relative and center-of-mass motions,
i.~e., $\Psi({\bf R}, {\bf r}) = \Phi({\bf r})\psi({\bf R})$.
Next, it is supposed that the relative motion wave function
$\Phi({\bf r})$ obeys the Schr\"{odinger} equation with the
Hamiltonian $H^{rel}$ which includes all terms of total
Hamiltonian, Eq. (\ref{hambas}), except the last one describing
the kinetic energy of the center-of-mass motion:
\begin{equation}
\left\{ H^{rel}-E({\bf R},\nabla_R) \right\}\Phi({\bf r }) = 0.
\end{equation}

Note that in contrary to the standard adiabatic
approach\cite{Adiabat} the relative motion Hamiltonian (see Eq.
(\ref{W1W2})) depends not only on the center-of-motion coordinate
$\mathbf{R}$ but also on the gradient $\nabla_{R}$. It does not
complicates, however, the effective Hamiltonian derivation, but
one should take into account that now the eigenvalue of the
relative motion equation, which usually plays a role of the
effective potential for the center-of-mass equation, is an
operator and gives the correction to the kinetic center-of-mass
operator as well. So, the effective center-of-mass Hamiltonian has
to be presented as follows
\begin{equation}\label{hamcm}
H^{CM} = -\frac{\hbar^2}{2 M} \nabla_{R}^2 + E({\bf R},\nabla_R).
\end{equation}

We solve the relative motion, Eq. (8), by means of a perturbation
technique in magnetic field strength powers restricting our
consideration by the second order terms. This assumption is valid
in weak field regime ($\hbar \omega^*_c < 2R_y^*$), where $R_y^* =
\mu e^4 / 2 \varepsilon^2 \hbar^2 $ is the effective Rydberg, and
$\omega^*_c = e B / \mu c$ is the cyclotron-resonance frequency.
This is the magnetic field regime relevant for experiments with
non-homogeneous magnetic fields~\cite{GeimPt,Dubonos,Kub} which
contrasts with experiments in homogeneous magnetic fields where
usually the high field regime $\hbar \omega^*_c \gg 2R_y^*$
is reached. Notice that $\hbar \omega^*_c = 2R_y^*$
corresponds to $B\approx 5\thinspace$T for GaAs.

Now taking into account the cylindrical symmetry of the zero order
Hamiltonian
\begin{equation}
H_0 = -\frac{\hbar^2}{2\mu}\nabla_{r}^2 - \frac{e^2}{\varepsilon
r}, \nonumber
\end{equation}
we find the zero order wave function
\begin{equation}
\Phi^{n,m}_0({\bf r }) =
A_{n,m}\left(\frac{2\lambda_{n}r}{a_B^*}\right)^{|m|}
\exp\left(im\varphi - \frac{\lambda_{n} r} {a_B^*}\right)
L^{2|m|}_{n+|m|-1}\left(\frac{2\lambda_{n}r}{a_B^*}\right),
\end{equation}
corresponding to the zero order eigenvalue
\begin{eqnarray}\label{eqE0}
E_0(n,m) = - \lambda^2_{n} R_y^*, \quad \lambda_{n} =
\left(n-1/2\right)^{-1},
\end{eqnarray}
where the symbols $n$ and $m$ stands for the radial and angular
relative motion quantum numbers, $a_B^*$ is the
effective Bohr radius, $A_{n,m}$ is the normalization constant
determined by $\langle\Phi^{n,m}_0({\bf r})|\Phi^{n,m}_0({\bf r })
\rangle=1$, and $L^{(\alpha)}_n(x)$
is the generalized Laguerre polynomial~\cite{Lag}.
Solving in addition the first order equation
\begin{equation}
\{H_0 - E_0(n,m)\}\Phi^{n,m}_1({\bf r })=\{E_1(n,m) -
W_1\}\Phi^{n,m}_0({\bf r }),
\end{equation}
with the orthogonality condition $\langle\Phi^{n,m}_1({\bf r})
|\Phi^{n,m}_0({\bf r })\rangle=0$,
we obtain the first and the second eigenvalue corrections
\begin{eqnarray}
E_1(n,m) &=& \langle\Phi^{n,m}_0({\bf r })|W_1|\Phi^{n,m}_0({\bf r
})\rangle = \frac{e \hbar}{2 \mu c} \xi B_z({\bf R}) m,  \\
E_2(n,m) &=& \langle\Phi^{n,m}_0({\bf r })|W_2|\Phi^{n,m}_0({\bf r
})\rangle + \langle\Phi^{n,m}_0({\bf r
})|[W_1-E_1(n,m)]|\Phi^{n,m}_1({\bf r })\rangle \nonumber \\ &=&
\beta^{n}_{m}\frac{e^2{a_B^*}^2}{8\mu c^2}B_z({\bf R})^2 +
\alpha^{n}_{m} \frac{e^2\hbar^2{a_B^*}^2}{2R_y^*
M^2c^2}\nabla_R\left\{ B_z({\bf R})^2 \nabla_R \right\}.
\label{finw2}
\end{eqnarray}

Here the symbols $\alpha^n_m$ and $\beta^n_m$ are numerical
constants which follow from the averages in Eq. (\ref{finw2}). For
a magnetic field perpendicular to the $xy$-plane we obtained the
following values: $\alpha^1_0=21/128$, $\beta^1_0=3/8$,
$\alpha^2_0=365/64$, $\beta^2_0=35/4$, and $\alpha^2_{\pm
1}=145/128$, $\beta^2_{\pm1}=11/4$.

Inserting the obtained eigenvalue corrections into Eq.
(\ref{hamcm}) we obtain the final expression for the effective
Hamiltonian
\begin{eqnarray} \label{EqEndRel}
H^{CM} &=& -\frac{\hbar^2}{2M}
\nabla_{R}\left\{1-\alpha^{n_r}_{m_r} \frac{e^2{a_B^*}^2}{M c^2
R_y^*}B_z({\bf R})^2 \right\} \nabla_R \nonumber \\
&-&\lambda_{n_r}^2 R_y^* + \frac{e\hbar}{2\mu c}\xi m_r B_z({\bf
R}) + \beta^{n_r}_{m_r}\frac{e^2{a_B^*}^2}{8\mu c^2}B_z({\bf
R})^2,
\end{eqnarray}
which describes the center-of-mass motion of the exciton with
relative motion quantum numbers $(n_r,m_r)$ in the non-homogeneous
magnetic fields.

\section{Cylindrical Symmetry}

From now on we
limit ourselves to cylindrical symmetric magnetic field profiles
$B_z({\bf R}) = B_z(R)$. Then, the exciton center-of-mass wave
function $\psi({\bf R})$ can be written as:

\begin{equation}
\psi ({\bf R}) = \frac{1}{\sqrt{2\pi }}\exp \left\{ im_{R}\phi
\right\} \psi (R),
\end{equation}
where $m_R$ is the quantum number corresponding to the exciton
center-of-mass angular momentum. Hamiltonian Eq. (\ref{EqEndRel})
then leads to the following radial Schr\"odinger equation:

\begin{equation}    \label{fineq}
\left\{- \frac{\hbar ^{2}}{2R}\frac{d}{dR}\left[
\frac{R}{M^{eff}(R)}\frac{d}{dR}\right] + V^{eff}(R)-E\right\}\psi
(R)=0, \label{finaleq}
\end{equation}
where
\begin{equation}
M^{eff}(R) = \frac{M}{1- \alpha^{n_r}_{m_r} c_1 B_z(R)^2},
 \label{effmass}
\end{equation}
is the effective mass for the center-of-mass motion of the
exciton, and
\begin{equation}
V^{eff}(R) = \beta^{n_r}_{m_r} c_2 B_z(R)^2 + m_r c_3 B_z(R) +
E_0({n_r}, {m_r}) + \frac{\hbar^2}{2M^{eff}(R)}
\frac{m_{R}^2}{R^2}, \label{effpot}
\end{equation}
is the effective potential for the exciton with
\begin{equation}
c_1 = \frac{e^2 {a_B^*}^2}{R_y^* M c^2} \qquad
; \qquad c_2 = \frac{e^2 {a_B^*}^2}{8\mu c^2} \qquad
; \qquad c_3 = \frac{e\hbar}{2\mu c}\xi.
\end{equation}

For GaAs we have typically $a_B^* \simeq 120\thinspace\AA$, $R_y^*
\simeq$ 5 meV, $c_1 \simeq 6.25 \times 10^{-3}\thinspace$
T$^{-2}$, $c_2 \simeq 0.44$ meV T$^{-2}$ and $c_3 \simeq 0.69$ meV
T$^{-1}$. Please note that the 2D exciton ground state energy
$E_0(1,0)$ is $-4R_y^* \simeq-20$ meV and the 2D exciton size is
one half the Bohr radius, $a_B^*/2 = 60 \thinspace \AA$. From the
above equations, we notice that the non-homogeneous magnetic field
modifies the exciton center-of-mass motion in the following ways.
First the exciton feels a $B_z(R)^2$ effective potential. It
implies that excitons will be collected in a region where the
magnetic field strength is minimum. Next, the exciton mass is
enhanced, it increases with $B_z(R)^2$, which favours the
localization of excitons. Third, the kinetic term gives a
contribution proportional to the gradient in the $B$-field, i.e.,
a $dB_z(R)/dR$ term. Finally, we see that the non-homogeneous
magnetic field also interacts with the exciton angular momentum,
and that this interaction is controled by the difference in mass
$\xi$. The reason is that the exciton is a neutral particle, and
if the electron and hole would have the same mass, the angular
momentum term would not contribute because, electron and hole give
the same contribution but with opposite sign.

\section{Magnetic Field Profiles}

We solve Eq. (\ref{fineq}) numerically  using the confinement
potential generated by two different non-homogeneous magnetic
field profiles, which experimentally are created by the deposition
of (a) a magnetized disk and (b) a superconducting disk on top of
a 2DEG, with a homogeneous magnetic field ($B_a$) applied
perpendicular to the 2DEG. Such an experimental configuration
results in a non-homogeneous magnetic field profile in the 2DEG.
The first profile (a) results into a magnetic dipole type of
profile, and the second (b) is also called the magnetic
antidot~\cite{BJ}. A sketch of the two experimental systems, the
magnetized disk and the superconducting disk, are shown in the
inset of Fig. 1(b) and Fig. 3(a), respectively.

\subsection{Magnetized Disk}

To calculate the magnetic field created by a magnetized disk, we
assume that the disk is very thin and is completely magnetized in
the z-direction. Therefore, we can write the magnetization in the
following way:

\begin{equation}
{\bf M}({\bf R})=h {\cal M} \delta(Z-d)\theta(a-R){\bf e}_Z,
\end{equation}
where $h$ is the disk thickness, $a$ is the disk radius, $d$ the
distance of the magnetic disk to the 2DEG, ${\cal M}$ the
magnetisation, $\theta(a-R)$ is the Heaviside step function, and
$Z$ and $R=\sqrt{X^2+Y^2}$ are cylindrical coordinates. The
corresponding vector potential can be calculated from the
differential form of Amp\'eres law $\nabla_R^2 {\mathbf
A}({\mathbf R})=-4\pi \nabla_R\times{\bf M}({\bf
R})$~\cite{Jackson}. In cylindrical coordinates, the vector
potential has only an angular component:

\begin{equation}
A_\varphi(R)=4B_0^D\sqrt{\frac{a}{R}}\frac{1}{p}
\left\{-E\left(p^2\right)+\left(1-\frac{p^2}{2}\right)K\left(p^2\right)\right\},
\end{equation}
with
\begin{equation}
p=2\frac{\sqrt{a R}}{\sqrt{\left(a+R\right)^2+d^2}},
\end{equation}
where $B_0^D = h {\cal M}$, and $K(x)$ ($E(x)$) is the elliptic
integral of first (second) type. The magnetic field can be
evaluated straightforwardly from ${\bf B}({\bf R}) = \nabla\times
{\bf A}({\bf R})$, which results into:

\begin{eqnarray}
B_z(R)&=&\frac{(a+R)}{(a+R)^2+d^2}A_\varphi(R) + B_0^D\frac{\left(
a^2-R^2+d^2 \right)}{R^2\sqrt{aR}}p^3 \nonumber
\\ && \times \left\{ -\frac{\partial }{\partial
p^2}E(p^2)-\frac{1}{2}K(p^2)+\left(1-\frac{p^2}{2}\right)
\frac{\partial }{\partial p^2}K(p^2)\right\}.
\end{eqnarray}

The magnetic field profile of the magnetized disk (the magnetic
dipole profile) in the 2DEG for $d=0.2 \thinspace \mu$m, $a=2
\thinspace \mu$m, and $B_0^D=0.1\thinspace$T is shown in the inset
of Fig. 1(a). Due to the quadratic dependence of the effective
potential on the magnetic field (see Eq. (22)), we can maximize
the confinement potential by applying a homogeneous field $B_a$ to
the magnetic dipole profile, $B_z^{total}(R)= B_{a} + B_z(R)$. We
use the experimental results of Dubonos {\it et al}~\cite{Dubonos}
in order to take values for the applied field ($B_a$) smaller than
the coercivity field, which has a strong dependence with the
radius of the magnetized disk~\cite{Dubonos}.

The magnetic dipole profile in the 2DEG for different values of
the relation $d/a$ are shown in Fig. 1(a) (Fig. 2(a)) as a
function of the radial coordinate, in the presence of a
homogeneous magnetic field $B_a=0.35\thinspace$T
($B_a=-0.25\thinspace$T), where we took $B_0^D=0.1\thinspace$T. We
also show in Fig. 1 ($B_a=0.35\thinspace$T) and in Fig. 2
($B_a=-0.25\thinspace$T) the effective potential, Eq.
(\ref{effpot}), for the exciton center-of-mass motion for the 1s
(Fig. 1(a) and Fig. 2(a)), and for the 2p$^{-}$ (Fig. 1(b) and
Fig. 2(b)) exciton relative motion quantum state, and the
effective mass (Fig. 1(b) and Fig. 2(b)), Eq. (\ref{effmass}), for
the 2p$^{-}$ exciton relative motion quantum state.

Notice that the effective potential, Figs. 1-2, is negative. This
is due to the energy shift resulting from the zero field exciton
relative motion $E_0(n_r,m_r)$ in the equation of the effective
potential (see Eq. (\ref{effpot})). This energy increases with
increasing principal relative motion quantum number $n_r$ (see Eq.
(\ref{eqE0})). Also notice that the effective potential of the
excited levels of the exciton relative motion (e.g., see Fig.
1-2(b) for the 2p$^-$ state) has a confinement region larger than
the 1s relative quantum level. The reason for this is as follows:
the coefficient of the diamagnetic term in the effective potential
equation (the term with quadratic dependence in the magnetic field
in Eq. (\ref{effpot})), $\beta^{n_r}_{m_r}$, is related to the
averaging of $r^2$ in the wave function of the exciton relative
motion, i.e., $\beta^{n_r}_{m_r}\propto
\langle\Phi^{n_r,m_r}_0({\bf r })|r^2|\Phi^{n_r,m_r}_0({\bf r
})\rangle$ (see Eq. (\ref{finw2})) which increases for increasing
$n_r$. Then, the confinement for the $n_r$ exciton relative motion
states should be stronger than the one for the $n_r - 1$ relative
state. Further, the exciton angular momentum term in the exciton
Hamiltonian is, in practice, bigger than the diamagnetic term.
This is because the effective potential for the exciton relative
motion quantum states with $m_r \neq 0$ has confinement energies
larger than the one for the relative levels with $m_r = 0$
(compare, e.g., the energy range in Fig. 1(a) and Fig. 1(b)).

The last term in our effective potential, Eq. (\ref{effpot}), is
related to the angular momentum of the exciton center-of-mass
motion, $m_R$. In that case, it contributes to the potential like
a centrifugal term ($m_R/R^2$), and leads to a peak near $R=0$
(see dotted curve in Fig. 1-2(b)). Also notice that the effective
mass is position dependent (see dashed curve in Fig. 1-2(b)). It
is larger than the normal exciton mass $M = 0.407$ although its
increase is small.

\subsection{Superconducting Disk}

The magnetic antidot was recently discussed by Reijniers {\it et
al}\cite{BJ}, where the properties of electrons in a two
dimensional system confined by such a magnetic antidot were
investigated. They found that, the magnetic field profile below a
very thin type-I superconducting disk is given by:

\begin{equation}
B_{z}(R) = B_a \left\{ 1+\frac{2}{\pi }\left[ \frac{a\zeta
}{\left( a^{2}+\zeta ^{2}\right) }-\arctan \left( \frac{a}{\zeta
}\right) \right] \right\}+\frac{2B_{a}}{\pi }\frac{a\zeta \left(
a^{2}-\eta ^{2}\right) }{\left( a^{2}+\zeta ^{2}\right) \left(
\zeta ^{2}+\eta ^{2}\right) },
\end{equation}
with
\begin{eqnarray}
\zeta^2 = \frac{1}{2} \left[ \sqrt{\left( R^{2}+d^{2}-a^{2}\right)
^{2}+4a^{2}d^{2}}+\left( R^{2}+d^{2}-a^{2}\right) \right],
\nonumber \\ \eta^2 = \frac{1}{2} \left[ \sqrt{\left(
R^{2}+d^{2}-a^{2}\right)^{2}+4a^{2}d^{2}}-\left(
R^{2}+d^{2}-a^{2}\right) \right],
\end{eqnarray}
where $a$ is the disk radius, and $d$ the distance of the
superconducting disk to the 2DEG. In Fig. 3(a) the magnetic field
profile resulting from the superconducting disk (magnetic
antidot), and the effective potential, Eq. (\ref{effpot}), for the
1s exciton relative motion quantum state is shown for $a=0.3
\thinspace \mu$m and for $d/a=0.1$ (dashed and dashed-dot curves)
and $d/a=0.01$ (solid and dotted curves). The effective mass, Eq.
(\ref{effmass}), and the effective potential for the 2p$^-$
exciton relative motion quantum state is plotted in Fig. 3(b) for
$d/a=0.1$. We took in Fig. 3(a) (Fig. 3(b)) an applied magnetic
field $B_a$ of 2 T (1 T). Notice that the magnetic field in the
2DEG is suppressed below the superconducting disk and an overshoot
is found near the edge of the disk which becomes smoother with
increasing distance, $d$, between the superconducting disk and the
2DEG. The exciton will prefer to localize below the disk because
$V^{eff}$ is smaller there. This is no longer true for the 2p$^-$
state (see Fig. 3(b)) where now $V^{eff}$ has a local minimum near
the edge of the disk where the magnetic field exhibits the largest
gradient. This occurs when the applied field $B_a$ is such that
the diamagnetic term (quadratic in the magnetic field) in the
effective potential, Eq. (\ref{effpot}), becomes comparable to the
angular momentum term (linear in the magnetic field).

\section{Numerical Results for the Exciton trapping Energy}

We have calculated the trapping energy and wave functions of an
exciton in a GaAs 2DEG in the presence of the above
non-homogeneous magnetic field profiles (magnetic dipole and
magnetic antidot profiles). The electron and hole mass and the
dieletric constant used in our calculations are\cite{Coo}
$m_e=0.067m_0$, $m_h=0.34m_0$, and $\varepsilon = 12.53$,
respectively. The problem of finding the eigenfunctions and
energies of Eq. (\ref{finaleq}) was solved by using a similar
numerical discretization technique as was used in Ref. 20.

We define the exciton trapping energy as the difference between
the exciton energy in the homogeneous applied field $B_a$ and the
energy of the exciton in the non-homogeneous magnetic field
profile for the same exciton state. Following this definition, a
positive exciton trapping energy implies that the exciton is
trapped in the effective potential created by the field
inhomogeneity. The influence of all the terms in Eqs. (21) and
(22) on the exciton trapping are extensively discussed. The
possibility of exciton trapping by using the confinement potential
created by the magnetic dipole, Figs. 1-2, and the magnetic
antidot, Fig. 3, are analyzed. In all the following figures, if
not explicitly stated otherwise, the disk radius and the distance
to the 2DEG used for the magnetized disk and for the
superconducting disk are $a = 2 \thinspace \mu$m, and $a = 0.3
\thinspace \mu$m, respectively, and we took the distance to the
2DEG equal to $d=0.1a$.

The trapping energy for the ground state of the exciton, for the
center-of-mass quantum numbers ($n_R,m_R$) $=$ (1,0) (dashed
curves) and (1,1) (triangles) are shown in Fig. 4 for the magnetic
dipole profile with a homogeneous applied field of (a) $B_a =
0.35\thinspace$T and (b) $B_a = -0.25\thinspace$T as a function of
the magnetisation of the disk $B_0^D$. Similar results are shown
in Fig. 5 for the magnetic antidot profile as a function of the
applied magnetic field $B_a$. Notice that for both profiles in low
fields, the trapping energy is negative (i.e., unbound state),
indicating that a large part of the center-of-mass exciton wave
function is extended into the magnetic barrier region. The
($n_R,m_R$) $=$ (1,1) is an excited state and requires slightly
larger magnetic fields to become bound. With increasing magnetic
field the exciton becomes more and more confined which increases
the trapping energy $E_B$. The magnetic disk in the presence of a
negative applied field (Fig. 4(b)) has a critical field where the
trapping energy starts to decrease with increasing field. This is
due to the competition between the magnetic field generated by the
disk $B_0^D$, and the applied field $B_a$. When the field of the
disk increases, the center region of the corresponding effective
potential (see solid curve in Fig. 2(a)) which has a peak
structure can become comparable to $V^{eff}$ at the edge of the
disk which for large $B_0^D$ (e.g. for ($n_R,m_R$) $=$ (1,0) in
Fig. 4(b) this occurs for $B_0^D > 0.06\thinspace$T) may lead to a
complete vanishing of the confinement region. One can increase the
trapping energy by applying a stronger homogeneous field $B_a$,
which is not trivial due to the limited coercivity field and for
sufficient large $B_a$ it will flip the magnetisation of the disk
and we end up into the situation of Fig. 1(a).

The angular momentum interaction with the magnetic field is
responsible for important effects in the trapping of atoms by
non-homogeneous magnetic fields~\cite{Wolf}. In two-dimensional
exciton systems, the momentum-field interaction is of importance
in a variety of exciton properties. The exciton trapping energies
for the seven lowest levels of the center-of-mass motion, for the
(a) 2p$^-$, (b) 2s, and (c) 2p$^+$, exciton relative motion
quantum states are shown in Fig. 6 (Fig. 7) for the magnetic
dipole profile with $B_a=0.35\thinspace$T ($B_a=-0.25\thinspace$T)
as a function of $B_0^D$, and in Fig. 8 for the magnetic antidot
as a function of $B_a$. In the case of the magnetic dipole
profile, the exciton in the 2p$^\pm$ states are much more confined
than the 2s state. The reason is that the angular momentum term in
Eq. (\ref{effpot}) gives a confinement potential stronger than the
diamagnetic term. Notice that in the case of a negative applied
field, due to this strong confinement of the 2p$^\pm$ states (Fig.
7), only the 2s state is affected by the competition between the
fields $B_0^D$ and $B_a$ as was discussed for Fig. 3. The trapping
energy of the superconducting disk profile also increases with
increasing magnetic field but for the 2p$^-$ exciton relative
motion quantum state (Fig. 8(a)) there is a local maximum in the
trapping energy. This energy starts to decrease after some field
($B_a=1\thinspace$T in Fig. 8(a)), which is exactly when the
quadratic term (positive) begins to be comparable to the angular
momentum term (negative) in Eq.~(\ref{effpot}) (see effective
potential in Fig. 3(b)).

Also notice that the exciton center-of-mass quantum levels with
non zero angular momentum ($m_R\neq0$) have only slightly higher
energy than the levels with $m_R=0$. The reason is because the
$m_R=0$ term in the effective potential, Eq. (\ref{effpot}), only
is responsible for a very small peak near $R = 0$ which gives a
very small contribution to the total effective potential. The
trapping energy in Figs. 6-8 are typically two orders of magnitude
larger than the one for the 1s state of the exciton relative
motion. In general, the trapping energy corresponding to the $n_r$
exciton relative quantum state resulting from the magnetic field
inhomogeneity should be bigger than the one corresponding to the
$n_r - 1$ quantum state. This can be easily understood from the
effective potential equation, Eq. (\ref{effpot}). The coefficient
of the diamagnetic term, $\beta^{n_r}_{m_r}$, increases with
increasing $n_r$ which increases the exciton trapping energy.

Figs. 9(a,b) and Fig. 10 show the ground-state radial wave
function of the exciton center-of-mass motion (see Eq.
(\ref{finaleq})) for the magnetic dipole profile for
$B_a=0.35\thinspace$T and $B_a=-0.25\thinspace$T, and for the
superconducting disk, respectively, as a function of the radial
coordinate $R$, for the quantum numbers of the relative motion
($n_r,m_r$) $=$ (1,0), for different values of $B_0^D$ (magnetized
disk) and $B_a$ (superconducting disk). Notice that with
increasing magnetic field the exciton becomes more localized. The
wave functions in Figs. 9(b) and 10 correspond to the effective
potential profiles of Figs. 2(a) and Fig. 3(a) which have the
minimum of the potential near the center of the disk, and that is
where the exciton is localized. In contrast to the case of Fig.
9(a) the effective potential (see Fig. 1(a)) has its minimum near
the outer edge of the disk where the exciton becomes localized. In
the latter case the center-of-mass wave function will have a ring
like structure.

In order to investigate the dependence of the exciton trapping
energy on the size of the magnetic field inhomogeneity, we
calculated the trapping energy for the 2s exciton relative motion
quantum state (i.e., ($n_r,m_r$) $=$ (2,0)), as a function of the
disk radius $a$. The results are shown in Figs. 11(a,b) for the
magnetic dipole profile with $B_0^D = 0.05\thinspace$T and a
homogeneous field of $B_a=0.35\thinspace$T and
$B_a=-0.25\thinspace$T, respectively, and in Fig. 12 for the
magnetic antidot profile with $B_0 = 0.5\thinspace$T, in each case
for the seven lowest quantum numbers of the center-of-mass motion.
In Fig. 11 (Fig. 12) we took the distance to the 2DEG $d=0.2
\thinspace \mu$m (0.03 \thinspace $\mu$m). The trapping energy
increases with disk radius but it saturates for large $a$. The
reason is that for sufficient large $a$ only the width of the
potential minimum increases but no longer the depth and
consequently the trapping energy becomes equal to the depth of the
minimum in $V^{eff}$.

\section{Conclusions}
In conclusion, we have developed a formalism to describe the
exciton motion in a two dimensional system, in the presence of a
non-homogeneous magnetic field. We assumed that the length scale
for the variation of the magnetic field is large as compared to
the exciton radius. We have theoretically calculated the exciton
trapping energies for the non-homogeneous magnetic fields created
by a magnetized disk (the magnetic dipole type profile) and by a
superconducting disk (the magnetic antidot). The results shown
than the trapping energy has a significant dependence on the
radius and profile of the non-homogeneous magnetic field, as well
as on the homogeneous applied background magnetic field $B_a$.
Further, the exciton angular momentum interaction with the
non-homogeneous magnetic field can be responsible for shifts as
big as a factor of 10 in comparison with the states with zero
angular momentum. In the present analysis we neglected the spin of
the electron and the hole which may have an important effect on
the trapping energy when the effective $g$-factor is substantially
different from zero. The analysis of the latter will be left for
future research.

\acknowledgments{This work was supported by the Inter-university
Micro-Electronics Center (IMEC, Leuven), the Flemish Science
Foundation (FWO-VI), and IUAP (Belgium). J. A. K. Freire was
supported by the Brazilian Ministry of Culture and Education
(MEC-CAPES) and F. M. Peeters is a research director with the
FWO-Vl. We acknowledge stimulating discussions with F. Pullizzi}

\begin{figure} \label{potmagPz}
\caption{(a) Effective potential for $m_R =0$ and ($n_{r},m_{r}$)
$=$ (1,0) and the magnetic field profile $B(R)$ of the magnetized
disk (inset of figure magnetic field profile in the 2DEG in the
absence of a background field) with a positive homogeneous applied
field $B_a=0.35\thinspace$T, as a function of the radial
coordinate $R$, for $d/a = 0.1$ with $V^{eff}(R)$ (solid) and
$B(R)$ (dotted), and $d/a = 0.2$ with $V^{eff}(R)$ (dashed-dotted)
and $B(R)$ (dashed). (b) Effective potential $V^{eff}(R)$ for
$n_{r} = 2$, $m_{r} = -1$ with: i) $m_R =0$ (solid) with
corresponding effective mass $M^{eff}(R)$ (dashed), and ii)
$V^{eff}(R)$ for $m_R =1$ (dotted) as a function of $R$, for $d/a
= 0.1$. We took $B_0^D = 0.1\thinspace$T and the disk radius $a =
2.0 \thinspace \mu$m.}
\end{figure}

\begin{figure} \label{potmagNz}
\caption{(a) Effective potential for $m_R =0$ and ($n_{r},m_{r}$)
$=$ (1,0) and the magnetic field profile $B(R)$ of the magnetized
disk with a negative homogeneous applied field
$B_a=-0.25\thinspace$T, as a function of the radial coordinate
$R$, for $d/a = 0.1$ with $V^{eff}(R)$ (solid) and $B(R)$
(dotted), and $d/a = 0.2$ with $V^{eff}(R)$ (dashed-dotted) and
$B(R)$ (dashed). (b) Effective potential $V^{eff}(R)$ for $n_{r} =
2$, $m_{r} = -1$ with: i) $m_R =0$ (solid) with corresponding
effective mass $M^{eff}(R)$ (dashed), and ii) $V^{eff}(R)$ for
$m_R =1$ (dotted) as a function of $R$, for $d/a = 0.1$. We took
$B_0^D = 0.1\thinspace$T and the radius $a = 2.0 \thinspace
\mu$m.}
\end{figure}

\begin{figure} \label{potsuper}
\caption{(a) Effective potential $V^{eff}(R)$ for $m_R =0$,
($n_{r},m_{r}$) $=$ (1,0)  and corresponding magnetic field
profile $B(R)$ of the superconducting disk, as a function of the
radial coordinate $R$, for $d/a = 0.1$ with $V^{eff}(R)$
(dashed-dot) and $B(R)$ (dashed), and $d/a = 0.01$ with
$V^{eff}(R)$ (solid) and $B(R)$ (dotted), and (b) effective
potential $V^{eff}(R)$ for $n_{r} = 2$, $m_{r} = -1$, $m_R =0$
(solid) with corresponding effective mass $M^{eff}(R)$ (dashed),
and $V^{eff}(R)$ for $m_R =1$ (dotted), as a function of $R$, for
$d/a = 0.1$. We took the radius $a = 0.3 \thinspace \mu$m and an
applied field $B_0$ of (a) 2 T and (b) 1 T.}
\end{figure}

\begin{figure} \label{nrmr10magPNz}
\caption{Exciton trapping energy, $E_{B}$, for ($n_{R},m_{R}$) $=$
(1,0) (dashed curve) and (1,1) (triangles), and ($n_{r},m_{r}$)
$=$ (1,0), for the magnetic dipole profile as a function of
$B_0^D$, for (a) $B_a=0.35\thinspace$T and (b)
$B_a=-0.25\thinspace$T, and for $a = 2 \thinspace \mu$m and
$d/a=0.1$.}
\end{figure}

\begin{figure} \label{nrmr10super}
\caption{Exciton trapping energy, $E_{B}$, for ($n_{R},m_{R}$) $=$
(1,0) (dashed curve) and (1,1) (triangles), and ($n_{r},m_{r}$)
$=$ (1,0), for the magnetic antidot profile as a function of
$B_a$, and for $d/a = 0.1$, with $a = 0.3 \thinspace \mu$m.}
\end{figure}

\begin{figure} \label{BindEmagPz}
\caption{Exciton trapping energy, $E_{B}$, as a function of
$B_0^D$, for the magnetic dipole profile for a homogeneous applied
field $B_a = 0.35\thinspace$T, for $a = 2 \thinspace \mu$m and
$d/a = 0.1$, and for $n_{r} = 2$, and (a) $m_{r} = -1$, (b) $m_{r}
= 0$, and (c) $m_{r} = +1$. The different curves correspond to the
different exciton center-of-mass levels ($n_{R},m_{R}$) $=$ (1,0)
dashed curve, (1,1) triangles, (2,0) dotted curve, (2,1) squares,
(3,0) solid curve, (3,1) circles, and (4,0) dashed-dotted curve.}
\end{figure}

\begin{figure} \label{BindEmagNz}
\caption{The same as in Fig. 6 but now for the magnetic dipole
profile with a homogeneous applied field of $B_a =
-0.25\thinspace$T.}
\end{figure}

\begin{figure} \label{BindEsuperz}
\caption{The same as in Fig. 6 but now as a function of $B_a$, for
the magnetic antidot profile, for $a = 0.3 \thinspace \mu$m and
$d/a = 0.1$.}
\end{figure}

\begin{figure} \label{wavemagPNz}
\caption{Exciton center-of-mass radial wave function $\psi(R)$ as
a function of the radial coordinate $R$, for the magnetic dipole
profile, for (a) $B_a = 0.35\thinspace$T and (b) $B_a =
-0.25\thinspace$T, for ($n_{R},m_R$) $=$ (1,0), ($n_{r},m_{r}$)
$=$ (1,0), for $a = 2 \thinspace \mu$m and $d/a = 0.1$, and for
$B_0^D = 0.02$ T (dotted), 0.06 T (solid), and 0.1 T (dashed).}
\end{figure}

\begin{figure} \label{wavesuper}
\caption{The same as in Fig. 9(a) but now for the magnetic antidot
profile, and for $a = 0.3 \thinspace \mu$m and $d/a = 0.1$, and
for $B_{a} = 1$ T (dotted), 1.5 T (solid), and 2.0 T (dashed).}
\end{figure}

\begin{figure} \label{BindEXamagPNz}
\caption{Exciton trapping energy, $E_{B}$, for ($n_{r},m_{r}$) $=$
(2,0), as a function of the disk radius $a$, for the magnetic
dipole profile for (a) $B_a = 0.35\thinspace$T and (b) $B_a =
-0.25\thinspace$T, for $d=0.2 \thinspace \mu$m and
$B_0^D=0.05\thinspace$T, and for center-of-mass quantum numbers
($n_{R},m_{R}$) $=$ (1,0) dashed curve, (1,1) triangles, (2,0)
dotted curve, (2,1) squares, (3,0) solid curve, (3,1) circles, and
(4,0) dashed-dotted curve.}
\end{figure}

\begin{figure} \label{BindEXasuper}
\caption{The same as in Fig. 11(a) but now for the magnetic
antidot profile, and for $d = 0.03 \thinspace \mu$m and $B_{a} =
0.5$ T.}
\end{figure}

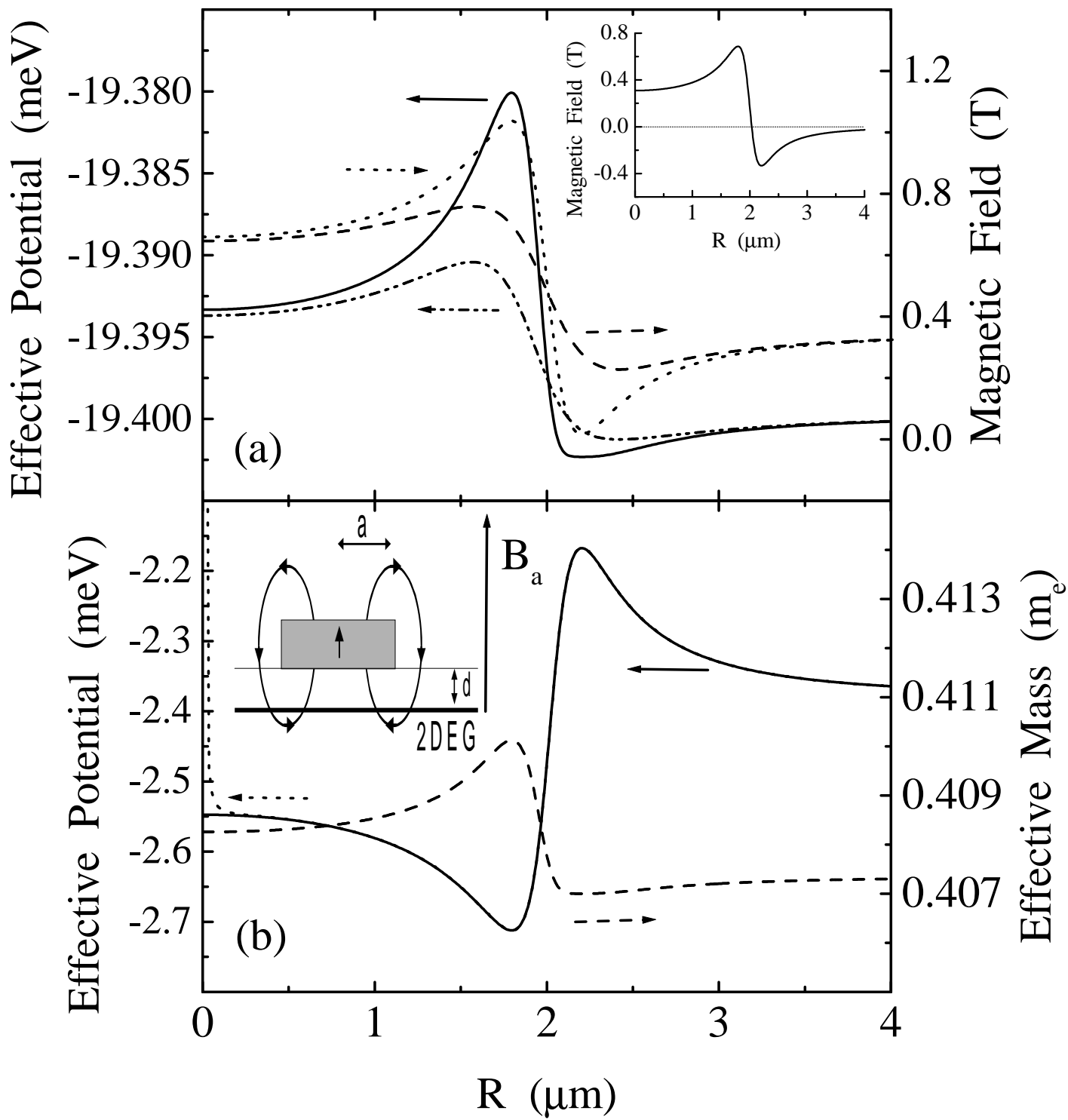
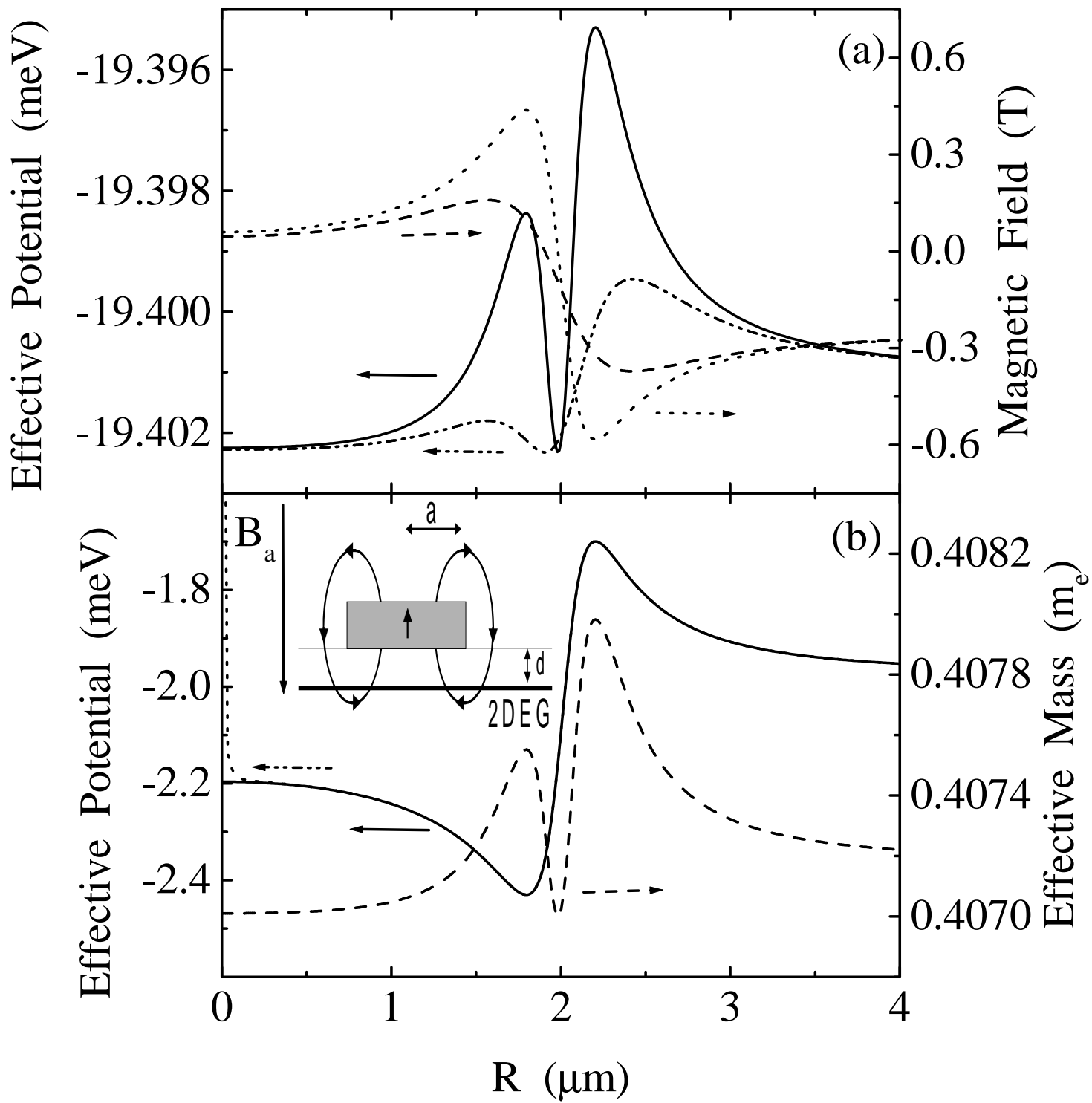
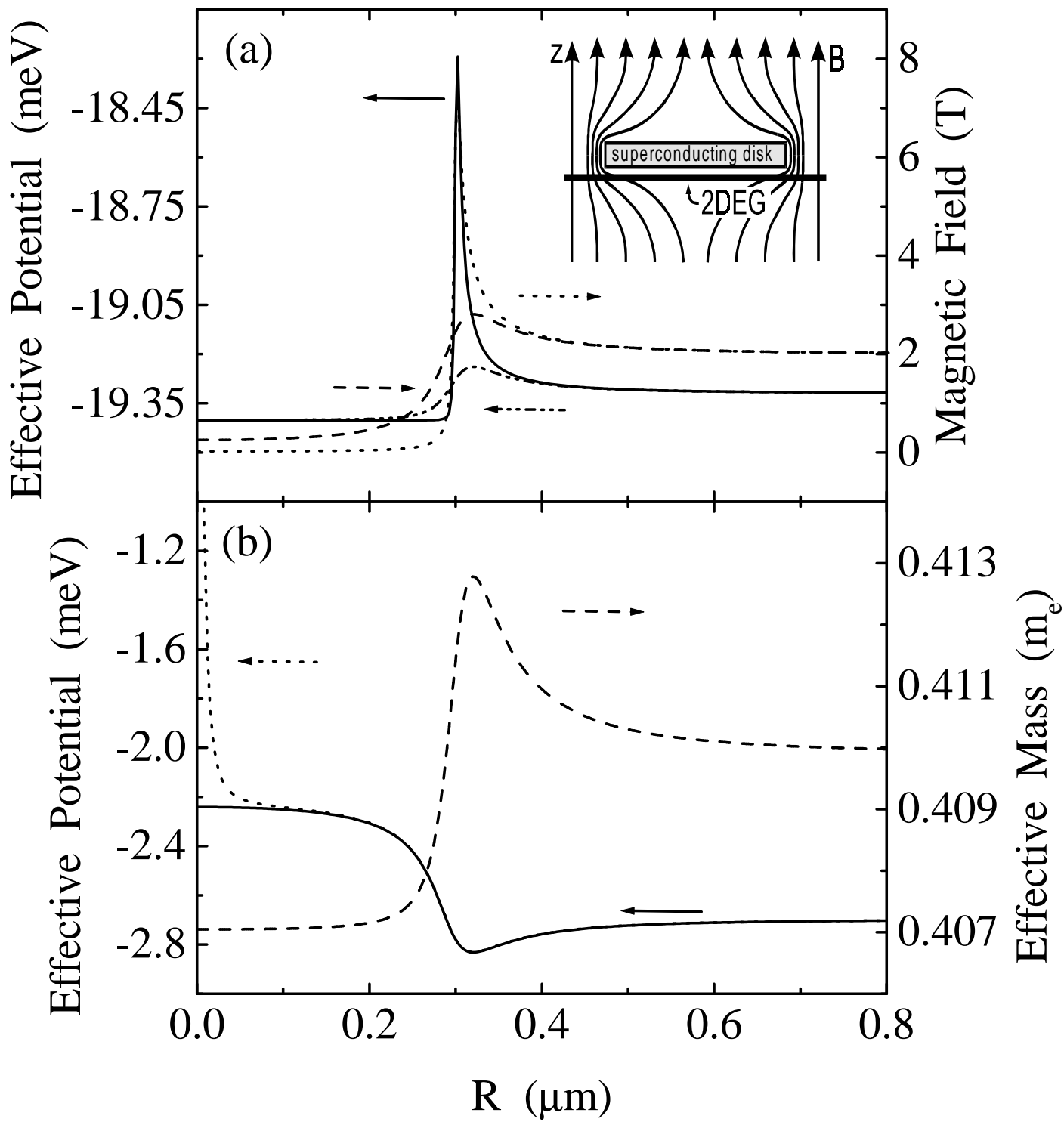
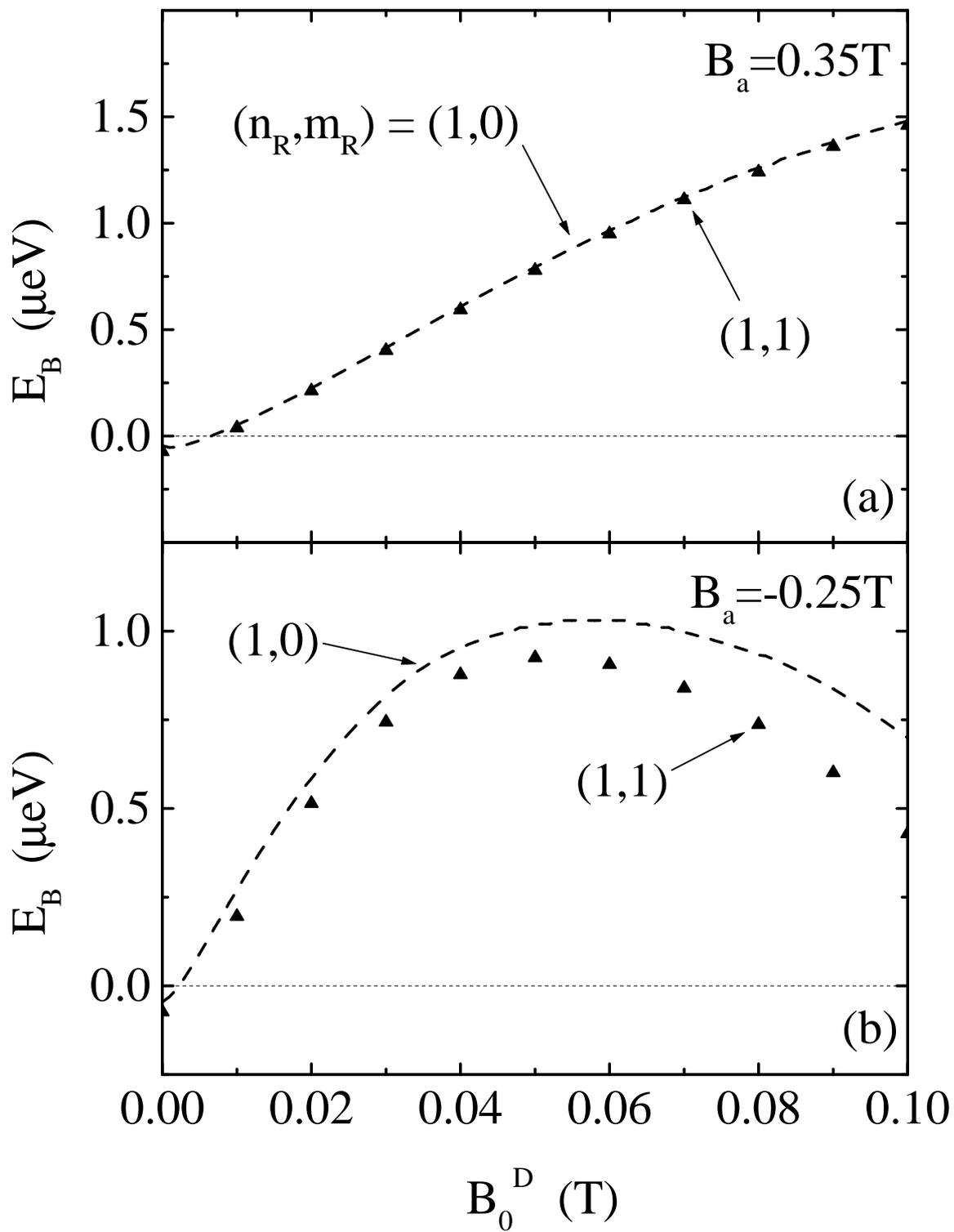
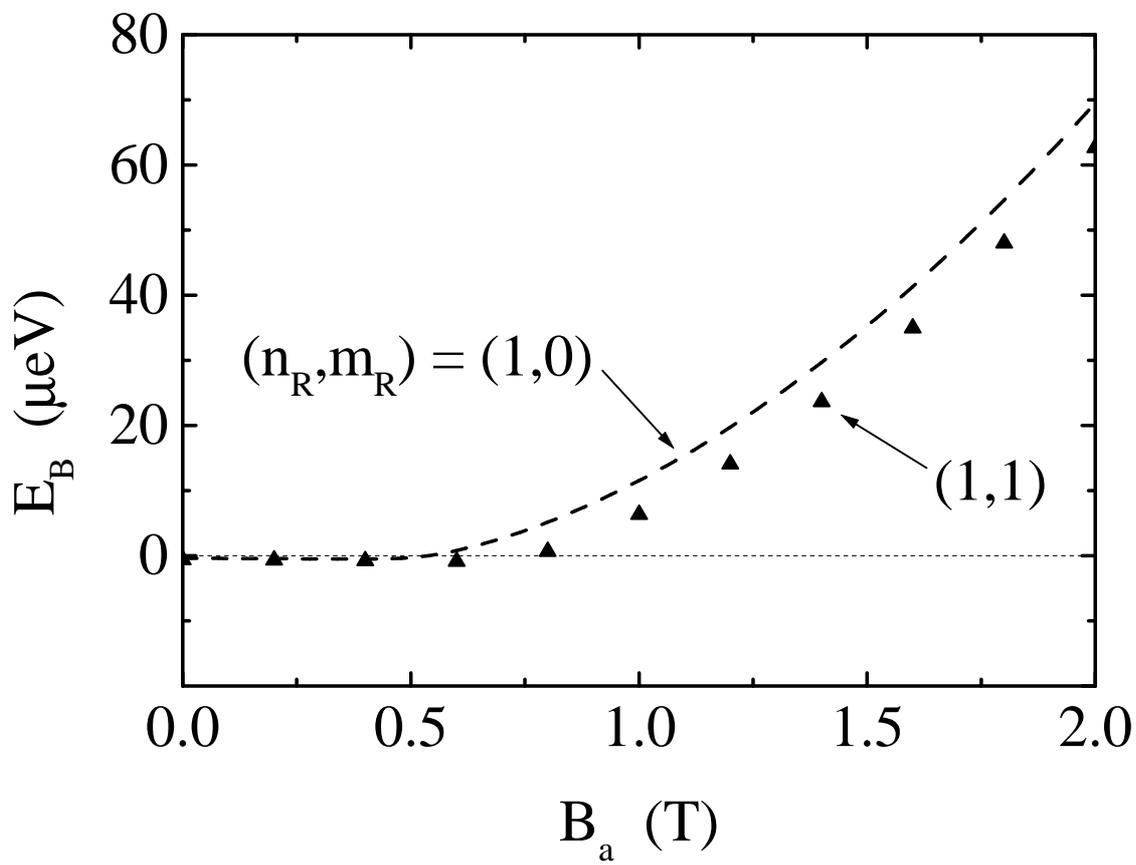
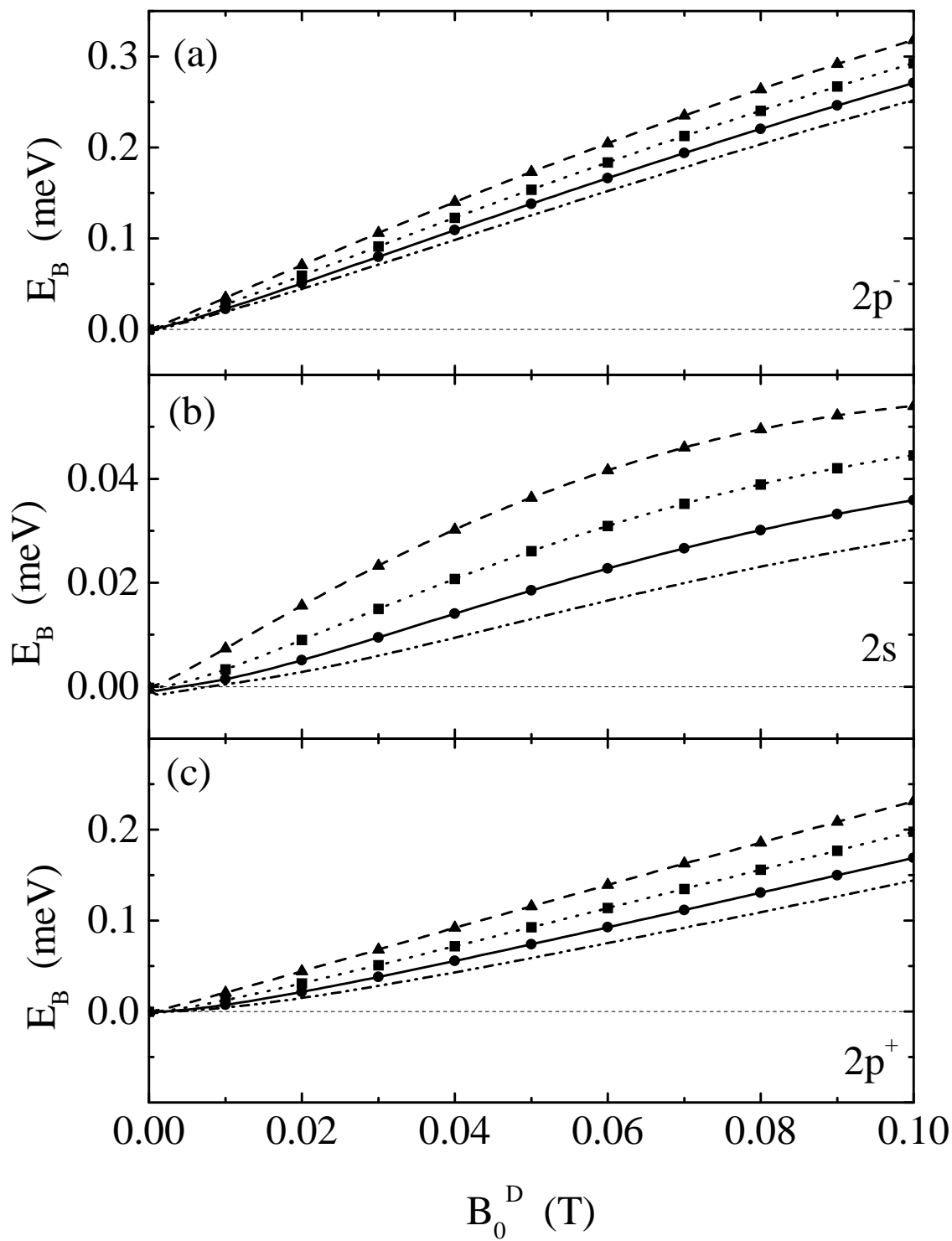
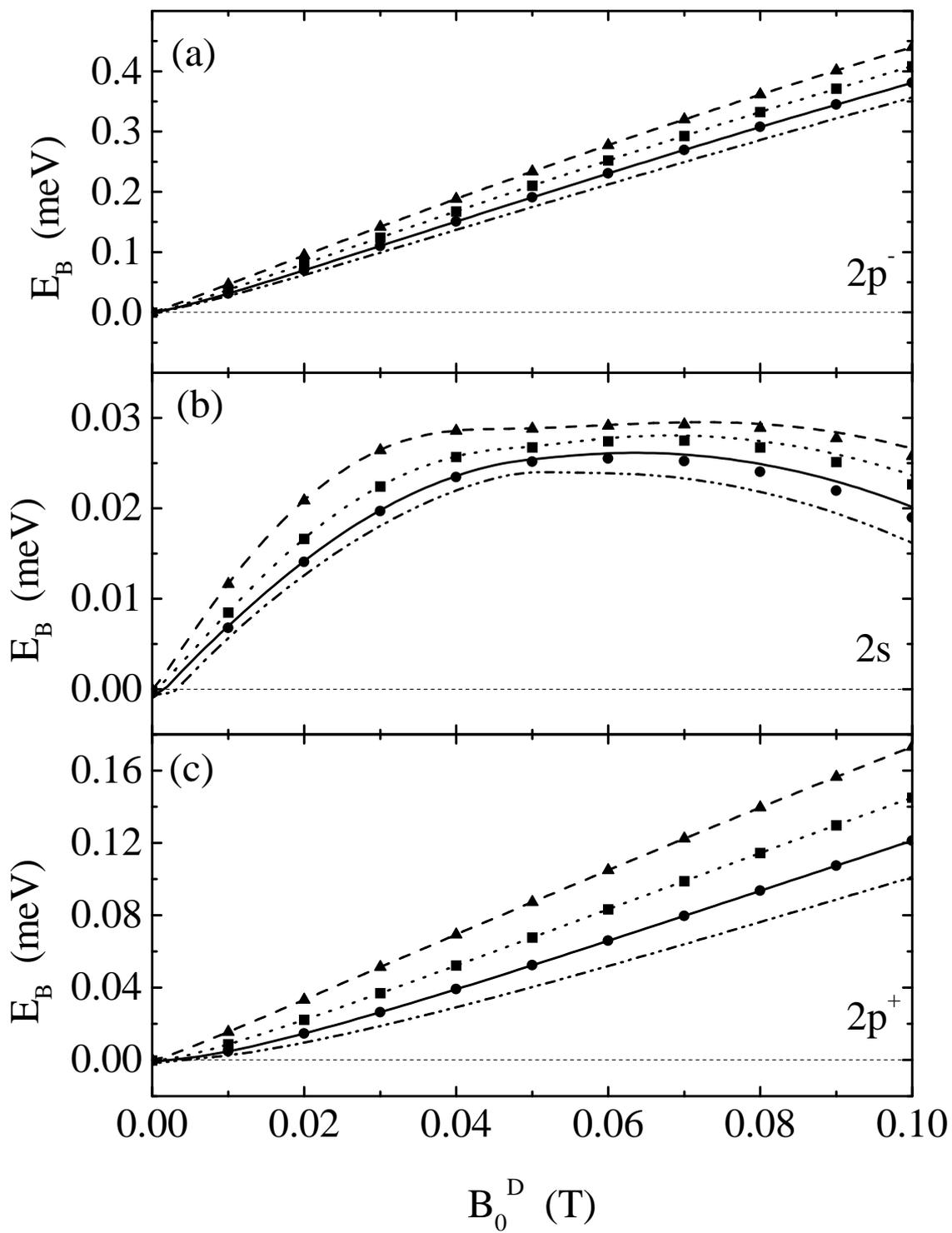
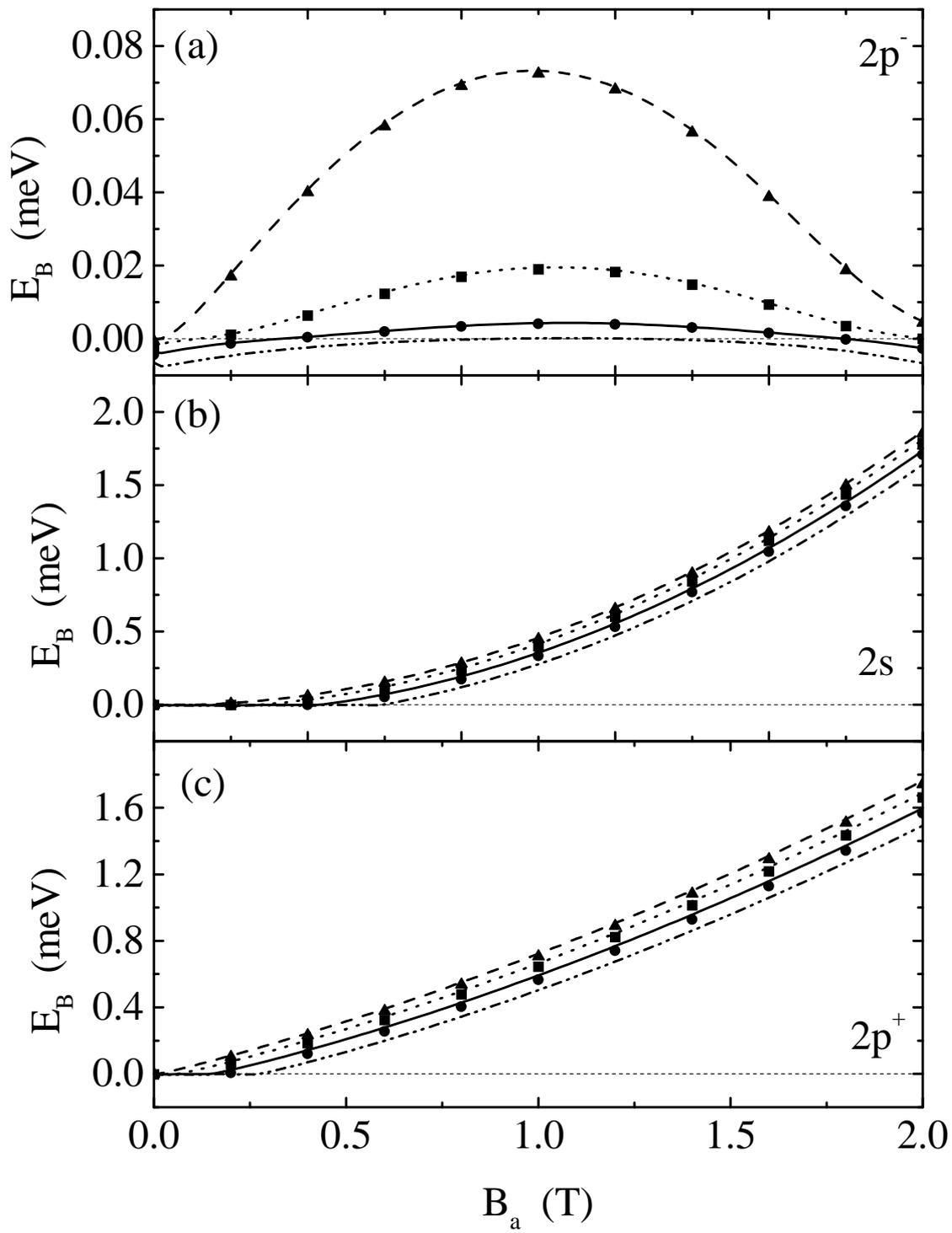
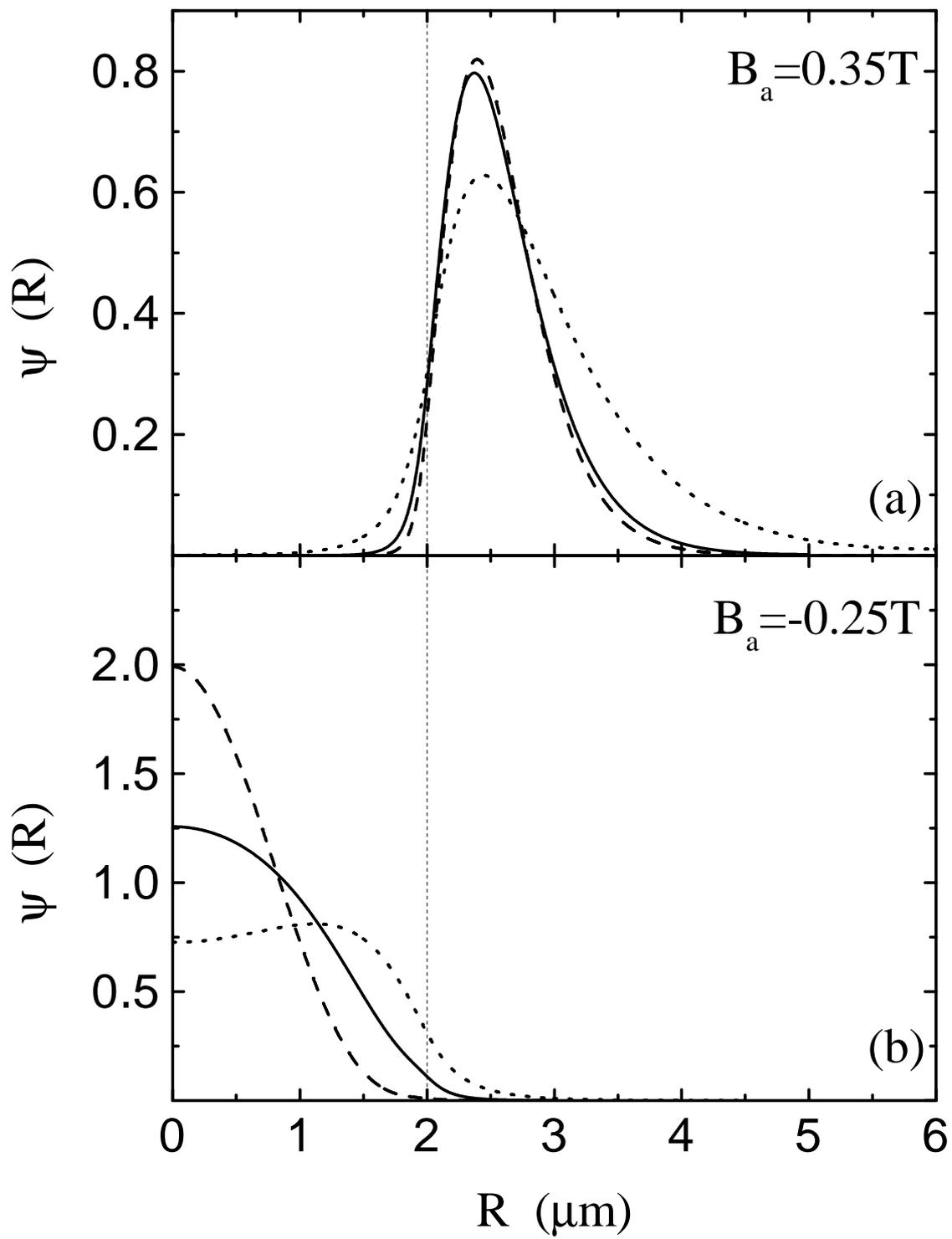
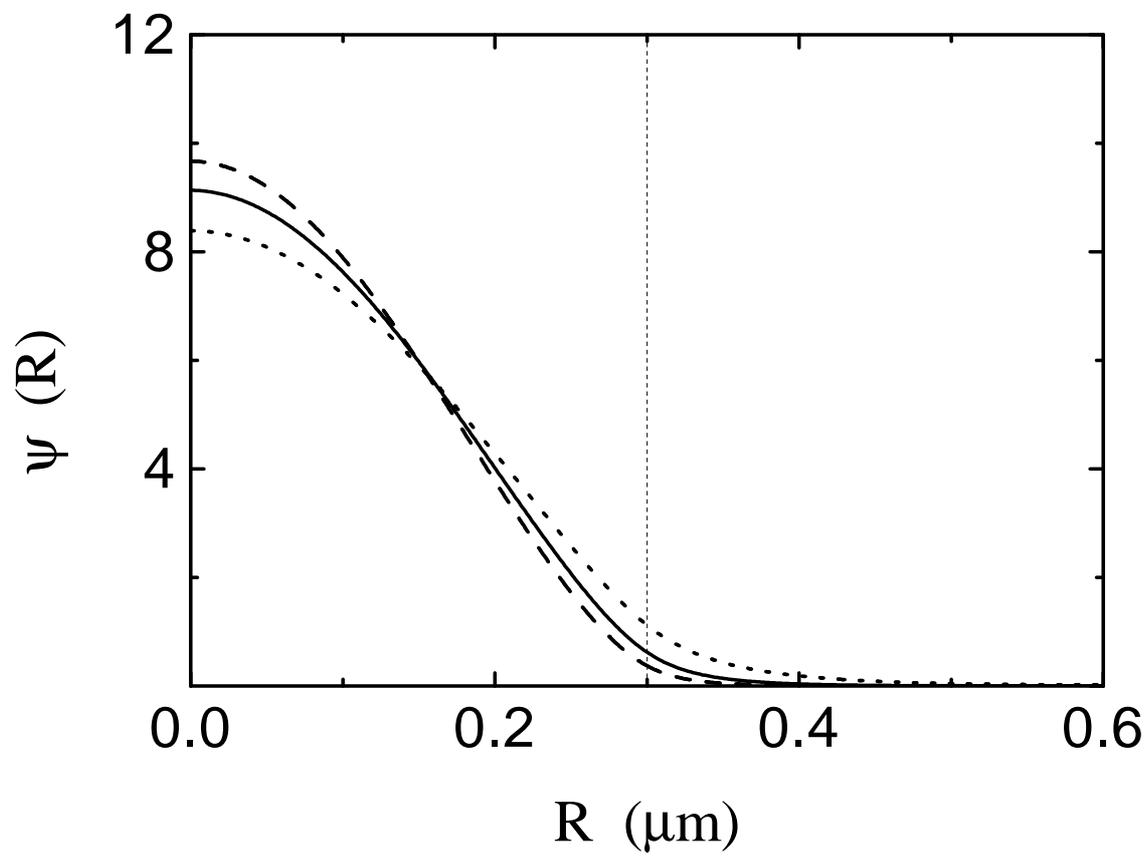
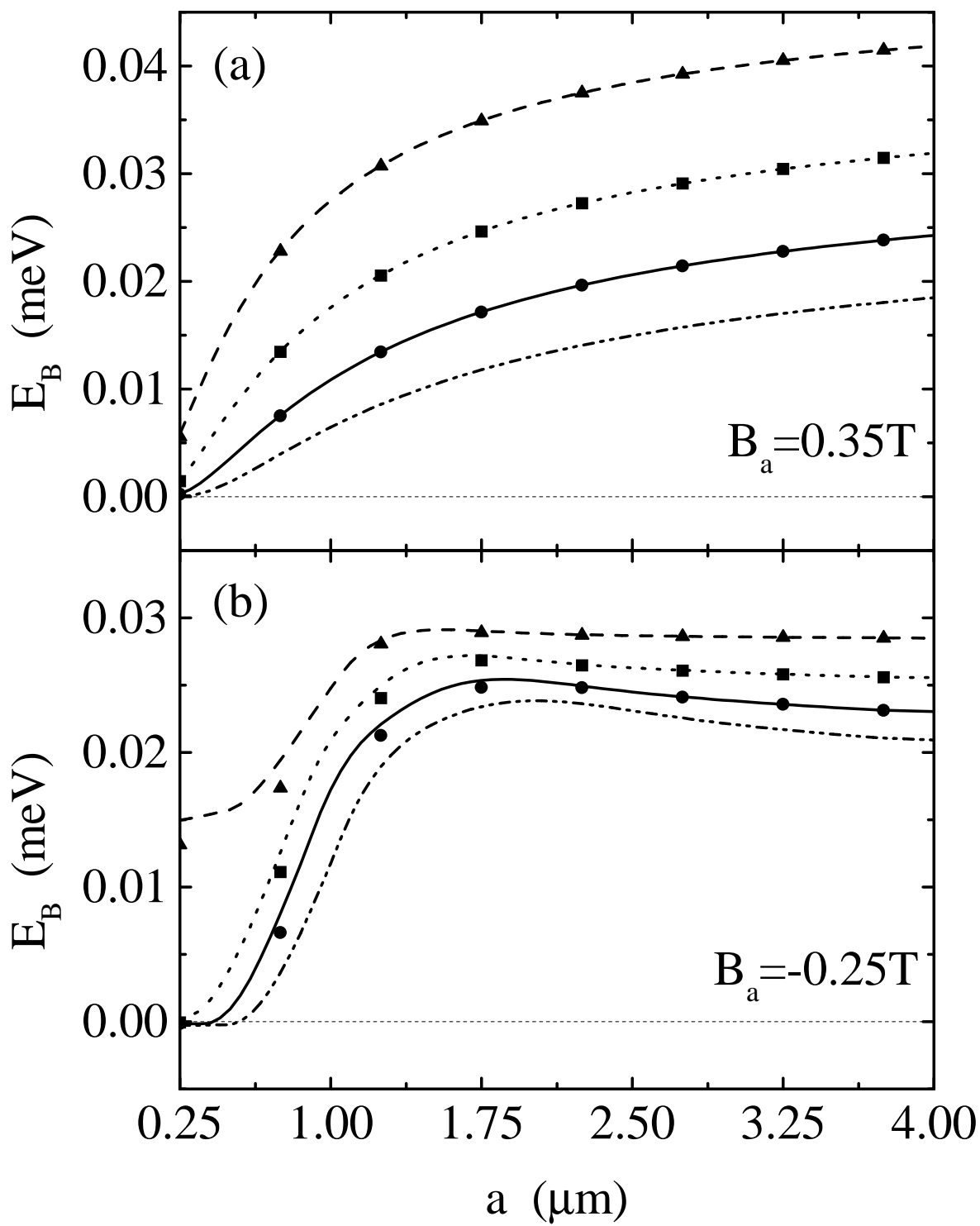
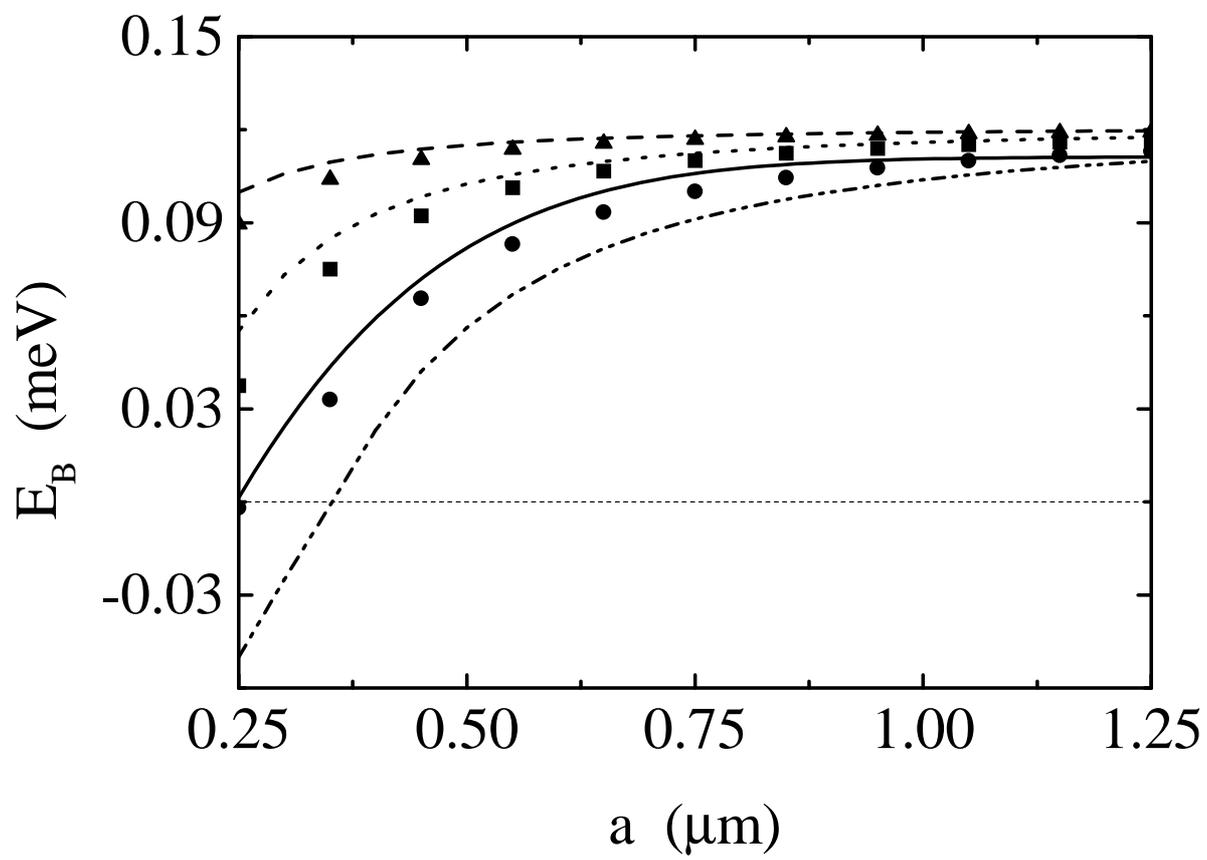
\end{document}